\documentclass[sigconf]{acmart}

\AtBeginDocument{%
  \providecommand\BibTeX{{%
    \normalfont B\kern-0.5em{\scshape i\kern-0.25em b}\kern-0.8em\TeX}}}
\setcopyright{acmcopyright}
\copyrightyear{2021}
\acmYear{2021}

\usepackage{verbatim} 
\usepackage{enumitem}

\begin{document}

\title{Personalized Embedding-based e-Commerce Recommendations at eBay}

\author{Tian Wang}
\email{twang5@ebay.com}
\affiliation{%
  \institution{eBay Inc.}
}

\author{Yuri M. Brovman}
\email{ybrovman@ebay.com}
\affiliation{%
  \institution{eBay Inc.}
}

\author{Sriganesh Madhvanath}
\email{smadhvanath@ebay.com}
\affiliation{%
  \institution{eBay Inc.}
}

\begin{abstract}
Recommender systems are an essential component of e-commerce marketplaces, helping consumers navigate massive amounts of inventory and find what they need or love. In this paper, we present an approach for generating personalized item recommendations in an e-commerce marketplace by learning to embed items and users in the same vector space. In order to alleviate the considerable cold-start problem present in large marketplaces, item and user embeddings are computed using content features and multi-modal onsite user activity respectively. Data ablation is incorporated into the offline model training process to improve the robustness of the production system. In offline evaluation using a dataset collected from eBay traffic, our approach was able to improve the Recall@k metric over the Recently-Viewed-Item (RVI) method. This approach to generating personalized recommendations has been launched to serve production traffic, and the corresponding scalable engineering architecture is also presented. Initial A/B test results show that compared to the current personalized recommendation module in production, the proposed method increases the surface rate by $\sim$6\% to generate recommendations for 90\% of listing page impressions. 
\end{abstract}


\maketitle

\section{Introduction}
\label{sec:intro}
Generating product recommendations for users is commonplace in e-commerce marketplaces. The eBay marketplace, with over 1.6 billion live items and over 183 million users, presents a unique set of challenges when it comes to generating recommendations. Traditional collaborative filtering and matrix factorization methods~\citep{recsys1} produce poor results given the scale and extreme sparsity of eBay's user-item matrix~\citep{galron2018deep}. With millions of new items listed daily, the cold start problem affects a substantial fraction of the inventory. Furthermore, over half of the live listings are single quantity, that is, they can be purchased by at most one buyer. After being purchased, items are removed from the site, and no longer accessible to users. Consequently, implicit user feedback signals such as clicks and purchases are extremely sparse. In this paper, we describe how we attempt to address these unique challenges to build an effective recommender system.

\begin{figure}
  \centering
  \includegraphics[width=\columnwidth]{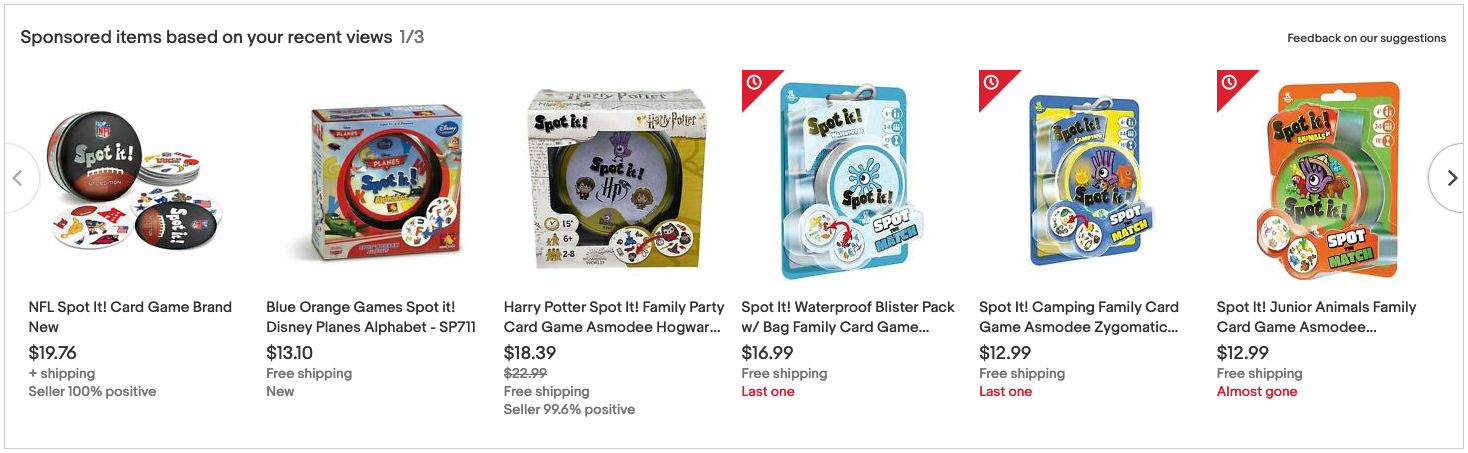}
  \caption{Screenshot of an eBay recommendations module where the user has been previously looking at games.}
  \label{fig:screenshot}
\end{figure}

Generally speaking, e-commerce recommendations may be driven purely by the shopping context or they may be personalized for a user based on a user profile. On an item listing page in an e-commerce marketplace, the $seed$ item provides strong indication of a user's shopping mission that may be used to guide the generation of recommendations. Indeed, there are several recommender systems based on the seed item context that are deployed at eBay~\cite{Brovman2, galron2018deep, Brovman1}. However, on other landing pages such as the homepage, such seed item context is missing. There may be other occasions as well where we need to provide personalized recommendations for the user in the absence of a seed item,  and the signal for generating recommendations is primarily available user information. Such "personalized recommendations" are our primary concern in this paper. Figure~\ref{fig:screenshot} depicts a screenshot of an eBay recommendations module ("sponsored items based on your recent views") where the input is primarily taken from a user's activity on the marketplace.

 Information about a user to generate a profile may be captured explicitly, by asking the user to fill out a survey as part of onsite registration, or implicitly, e.g. by parsing the user's shopping history. Although explicit methods directly capture the user's interests, there are several limitations with this approach: an exhaustive set of potential interests is difficult to curate, user participation tends to be low, the input can be highly incomplete, and long-term interests may not capture specific short-term shopping missions. Due to these limitations, generating a user profile of interests is commonly performed using implicit user interaction data. 

In this paper, we propose to model users as embeddings based on implicitly observed user shopping behavior. Using a two-tower deep learning model architecture~\cite{covington2016deep}, one tower for items and one for users, users and items are represented as points in the same vector space. In order to address the data sparsity and cold-start challenges mentioned above, (i) items are represented using content-features only, and (ii) we expand the set of implicit user signals to incorporate multi-modal user onsite behaviors such as item clicking and query searching. Once trained, a k-nearest neighbor (KNN) search using a user embedding is used to generate a set of item recommendations for the user that reflect his or her implicit shopping behavior. At runtime, an additional Learning-To-Rank (LTR) model may be applied to this candidate item set in order to improve conversion, as was done in the work by ~\citet{Brovman2}. However this paper primarily focuses on the method for generating personalized recommendation candidate items. And since deploying a deep learning based recommendation model to a large scale dynamic industrial marketplace environment involves non-trivial engineering challenges, we also discuss details of our production engineering architecture. In summary, we contribute methods and techniques for:
\begin{enumerate}[label={(\roman*)}]
 \item generating content-based item embeddings to address the cold-start problem
 \item generating multi-modal user embeddings from various onsite events, such as item views and search queries
 \item selectively dropping out training data to increase production model robustness
 \item utilizing cluster-based KNN algorithm to increase recommended item diversity
 \item deployment of the model and end-to-end recommender system to eBay's large scale industrial production setting
\end{enumerate} 

This paper is organized in the following manner. Section~\ref{sec:related} summarizes related work from academia as well as industry. We describe the proposed core model architecture in Section~\ref{sec:model}. The dataset as well as the offline experiments to evaluate the model are presented in Section~\ref{sec:data}. To analyze the model robustness in production environment, we conduct user data ablation analysis, and propose solutions to improve model performance. We then turn our attention to the model prediction stage in Section~\ref{sec:eng}, cover retrieval as well as the production engineering architecture and discuss empirical A/B test results. Finally, we present a summary of this work and discuss future directions in Section~\ref{sec:conc}.

\section{Related Work}
\label{sec:related}
The generation of personalized recommendations is a well studied problem in both academia and industry. Among the most popular techniques are matrix factorization models (e.g. \citep{hu2008collaborative, koren2009matrix, rendle2009bpr}) which decompose a user--item matrix into user and item matrices, and treat recommendation as a matrix imputation problem. Despite seeing success in the Netflix competition for movie recommendation \citep{koren2009matrix}, traditional matrix factorization models require unique user and item identifiers, and do not perform as well in a dynamic \mbox{e-commerce} marketplace where existing items sell out and new items come in continuously. Utilizing content features such as the item title text becomes essential for tackling data sparsity and cold-start issues, and various methods have been proposed to address this within the matrix factorization framework. For example, Content-boosted collaborative filtering \citep{melville2002content} uses a content-based model to create pseudo user-item ratings. Factorization machine \citep{rendle2010factorization} and SVDFeature \citep{chen2012svdfeature} directly incorporate user and item features into the model. 

More recently, neural networks have been used to model more complex content features and combine them in a non-linear fashion. \citet{covington2016deep} proposed two-tower neural networks to embed users and items separately, and applied it to the task of generating video recommendations. \citet{he2017neural} explored the use of a non-linear affinity function to replace the dot product between the user and item embedding layers for improved model capacity. \citet{zhu2018learning} and \citet{gao2020deep} further extended the idea by using graph structures for candidate recall and scaling the non-linear affinity function for an industrial setting, for e-commerce and video recommendations respectively. Our work takes inspiration from these efforts and the practical challenges and limitations posed by the eBay marketplace. 

There is a different but related line of work focusing on using neural networks for LTR, such as Deep and Wide \citep{cheng2016wide} and DIN \citep{zhou2018deep}. However, our work is is aimed at tackling the core candidate recall retrieval problem in an industrial setting, with the primary goal of efficiently selecting small groups of relevant items from a very large pool of candidate items. As mentioned earlier, an LTR model may be applied to this candidate item set to improve user engagement and conversion. 

\section{Model}
\label{sec:model}
\begin{figure*}[t!]
	\small
    \centering
    \begin{minipage}{0.95\textwidth}
    \includegraphics[width=\columnwidth]{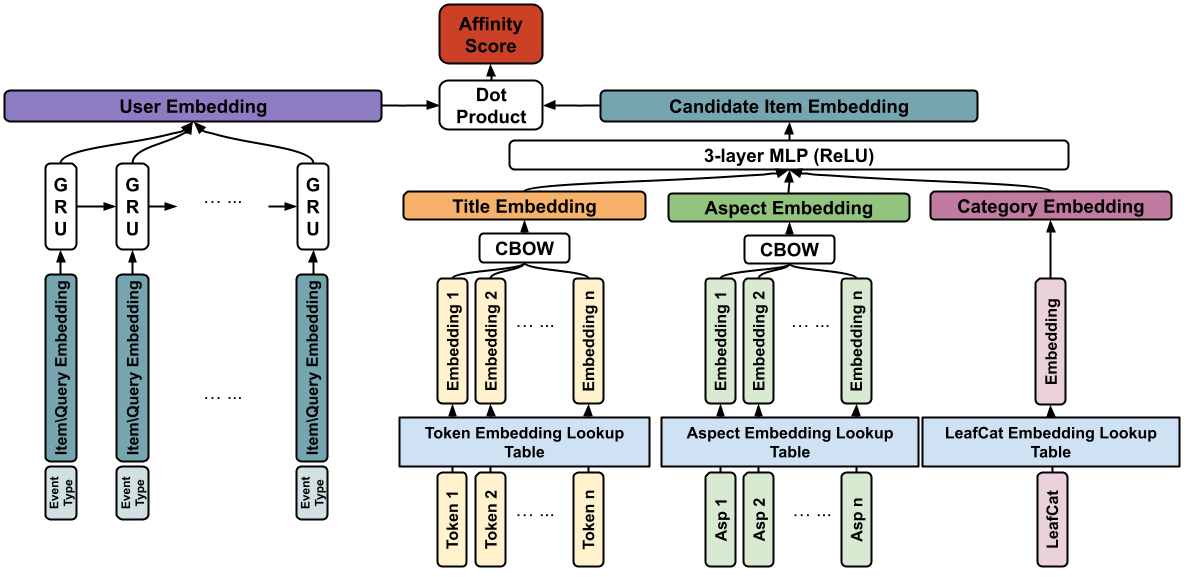}
    \end{minipage}
    \hfill
    \caption{Model architecture with recurrent user representation.}
    \label{model}
\end{figure*}
Our proposed approach for personalized recommendations is based on training a two-tower deep learning model to generate user embeddings and item embeddings at the same time. The architecture of the model is as shown in Fig.~\ref{model}, and described in detail below. We also mention the impact of adding specific model features to our primary offline model performance metric, Recall@K, described in detail in Section~\ref{sec:metrics}.

Following the work by \citet{covington2016deep}, we model generating recommendations as a classification problem with the softmax probability:

\begin{equation}
    P(s_i | U)  = \frac{e^{\gamma(\mathbf{v}_i, \mathbf{u})}}{\sum_{j \in V} e^{\gamma(\mathbf{v}_j, \mathbf{u})}},
\label{eq:classification}
\end{equation}
where $\mathbf{u}\in \mathbb{R}^D$ is a $D$-dimensional vector for the embedding of user $U$, $\mathbf{v}_i\in \mathbb{R}^D$ is a $D$-dimensional vector for the embedding of item $s_i$, $\gamma$ is the affinity function between user and item, and $V$ is all items available on eBay. As $V$ could contain billions of items, it is infeasible to perform a full-size softmax operation. Negative sampling has to be used to limit the size of $V$, and we will discuss this further in Sec. \ref{sec:negative}. The whole model is trained to minimize the negative log-likelihood (NLL) of observed user clicks in the dataset. Next, we discuss the details of how eBay items are encoded by the model.

\subsection{Content-based Item Embedding}\label{item_embedding}

In the eBay marketplace, an item corresponds to a listing (or offer) of something for sale from a seller. In order to address the cold-start problem, an item in our model is represented not as a unique identifier (item id), but solely by using its content-based features such as item title, category (e.g mobile phone), and structured aspects (e.g brand: Apple, network: Verizon, etc.). We chose not to incorporate historical item-behavior features (e.g. historical Click-Through-Rate, Purchase-Through-Rate) in our model. These features are not applicable to cold-start items and are constantly changing by their very nature, creating additional engineering complexity for storage and retrieval when building a large-scale production system. 

For title and aspect features, we tokenize and convert raw text into token embeddings with embedding size $D_{text}$, and use the Continuous-Bag-of-Words (CBOW) \citep{mikolov2013efficient} approach to generate title and aspect feature representations. The vocabulary for the title feature consists of approximately 400K tokens and is gathered from eBay item titles as opposed to a generic English language corpus. This allows us to better capture the distribution of item title tokens in the eBay marketplace, which is drastically different from the traditional English language, as is demonstrated in the work by \citet{wang2020item}. Tokenization is comprised of replacing any character that is not \textit{a-Z} or a number with whitespace, and splitting by whitespace. The vocabulary for aspect features comes from the existing production database and contains around 100K aspect tokens. For the item category feature, we index the category values and map them into an embedding space of size $D_{category}$ using a lookup table. All of the embedding tables are trained from scratch with random initialization from the standard normal distribution $\mathcal{N}(0, 1)$.

After mapping all item features into a continuous space, item feature embeddings $\mathbf{z}_i$ are concatenated and passed through a MLP with $L$ hidden layers, $H$ hidden dimensions, and Rectified Linear Units (ReLU)~\cite{glorot2011deep} as the non-linear activation function, to generate a $D$-dimensional item embedding $\mathbf{v}_i$: 
\begin{equation}
\begin{split}
    \mathbf{z}_i & = \text{concat}(\mathbf{z}^{\text{title}}_i, \mathbf{z}^{\text{aspect}}_i, \mathbf{z}^{\text{category}}_i), \\
    \tilde{\mathbf{v}_i} & = \text{MLP}(\mathbf{z}_i). \\
    \mathbf{v}_i & = \frac{\tilde{\mathbf{v}_i}}{||\tilde{\mathbf{v}_i}||}
\end{split}
\end{equation}
The item embedding $\mathbf{v}_i$ is normalized to unit length. We now turn our attention to the user tower part of the model.

\subsection{Multi-Modal User Embedding}

A user's activity on an e-commerce marketplace is not limited to only viewing items. A user may also perform actions such as making a search query, adding an item to their shopping cart, adding an item to their watch list, and so on. These actions provide valuable signals for the generation of personalized recommendations. In this work, we have attempted to create a generic framework to incorporate such "multi-modal" user activity into the model. We have chosen to start with item viewing and the search query user actions as representatives of item-based events and query-based events respectively, since these are the quintessential online shopping activities. 

Item views/clicks are the most common form of implicit user feedback for an e-commerce marketplace, and generate large volumes of training data. For an item-based event $z_i$, we first map the corresponding item $s_{z_i}$ to the corresponding embedding $\mathbf{v}_{z_i}$ as described in Sec. \ref{item_embedding}, and then concatenate it with a 4-dimensional vector $\mathbf{e}_{z_i}$ representing its event type. 

User searches are a valuable signal for a recommender system as they are strong indications of explicit user interest or shopping mission. In order to encode this user action into our framework, we model each search query as a \textit{"pseudoitem"} with the actual query text taking place of the item title, and the "dominant" query category (predicted using a separate model) taking place of the item category, and the aspects left empty. The event type embedding is concatenated to the item-based embedding. Adding this search query signal to the model resulted in a $\sim$4\% improvement in our offline validation metric, Recall@20.

We denote for each user event $z_i$, its corresponding vector representation $E(z_i)$ as:
\begin{equation}
    E(z_i) = \text{concat}(\mathbf{v}_{z_i}, \mathbf{e}_{z_i})
\end{equation}

We explored different methods of generating a user embedding for a given user $U$ with onsite activity $Z = \{z_1, ..., z_n\}$.

\subsubsection{Continuous Bag-of-Events Representation}
The first approach is to bag all the event embeddings into a single vector by averaging over all embeddings. After combining all events into a single vector, we use a MLP layer with $L$ layers, $H$ hidden dimension, and ReLU non-linear activation functions to generate a $D$-dimensional user embedding $\mathbf{u}$: 
\begin{equation}
\begin{split}
    \mathbf{\tilde{u}} &= \text{MLP}(\frac{1}{n} \sum_{i=1}^n E(z_i)),\\
    \mathbf{u} &= \frac{\mathbf{\tilde{u}}}{||\mathbf{\tilde{u}}||} \\
\end{split}
\end{equation}
Continuous Bag-of-Events is the simplest representation of user activity, however, in this approach, the ordering of the events does not affect the outcome. 

\subsubsection{Recurrent Representation}
In order to integrate the ordering information of user historical events, we also experimented with using a recurrent neural network to process the sequence of event embeddings. We start with gated recurrent units \citep[GRU,][]{cho2014learning}, which have the update rule $\mathbf{h}_t = \phi (\mathbf{x}_t, \mathbf{h}_{t-1})$ defined by:
\begin{equation}
\label{eq:gru}
\begin{split}
\mathbf{r}_t & = \mathbf{\sigma}  (\mathbf{W}_r \mathbf{x}_t + \mathbf{U}_r \mathbf{h}_{t-1}) \\
\mathbf{u}_t & = \mathbf{\sigma}  (\mathbf{W}_u \mathbf{x}_t + \mathbf{U}_u (\mathbf{r}_t \odot \mathbf{h}_{t-1})) \\
\tilde{\mathbf{h}}_t & = \tanh(\mathbf{W} \mathbf{x}_t + \mathbf{U} (\mathbf{r}_t \odot \mathbf{h}_{t-1})) \\
\mathbf{h}_t & = (1 - \mathbf{u}_t) \odot \mathbf{h}_{t-1} + \mathbf{u}_t \odot \tilde{\mathbf{h}}_t,
\end{split}
\end{equation}
where $\sigma$ is a sigmoid function, $\mathbf{x}_t$ is the input at the $t$-th timestep, and $\odot$ is element-wise multiplication. 

We initialize the GRU recurrent hidden state $\mathbf{l}_0$ as 0. For each event $z_t$ in the user history $Z$, we feed the corresponding event embedding into the GRU cell in sequence as input in each timestep:
\begin{equation}
\begin{split}
    \mathbf{l}_t & = \phi (E(z_i), \mathbf{l}_{t-1}), \\
    \mathbf{l}_0 & = \mathbf{0}. \\
\end{split}
\end{equation}

The $D$-dim user embedding $\mathbf{u}$ is generated by taking the average over output vectors from all GRU steps:
\begin{equation}
\begin{split}
    \mathbf{\tilde{u}} & = \frac{\sum_{i=1}^n \mathbf{l}_i}{n},\\
    \mathbf{u} & = \frac{\mathbf{\tilde{u}}}{||\mathbf{\tilde{u}}||} \\
\end{split}
\end{equation}

Compared to the Continuous Bag-of-Events user representation, recurrent user representation has access to the order of user activity, and in principle can better relate user relevance feedback to the user's interaction history. In our experiments, using this recurrent user representation in our model resulted in a $\sim$5\% gain in our offline Recall@20 metric. 

\subsection{Affinity function}

The affinity function $\gamma(\mathbf{v}_i, \mathbf{u})$ between user $U$ and item $s_i$ is constructed by the dot product between the user and item embeddings. As user and item embeddings are normalized to have unit length ($||\mathbf{u}|| = 1$, $||\mathbf{v}_i|| = 1$), the dot product score between any pair of embeddings is constrained to have a value between -1 and 1. This essentially limits the capability of the model to distinguish positive items from negatives items. In order to address this, we added a \textit{temperature} $\tau$ \citep{yi2019sampling} term to our affinity function described in Eq.~\ref{eq:classification} as follows:
\begin{equation}
\gamma(\mathbf{v}_i, \mathbf{u}) = \frac{\mathbf{v}_i\mathbf{u}}{\tau}.
\end{equation}
The temperature hyperparameter was tuned to maximize the retrieval metric, Recall@k. In our experiments, we found that $\tau$ has a large impact on the performance of the trained model. By tuning $\tau$ on the validation set, we are able to increase Recall@20 by $\sim$150\%. 

\section{Dataset \& Experiments}
\label{sec:data}
In this section, we describe the dataset we created to train our model, the importance of negative sampling during the training process, as well as the offline experiments performed to evaluate the effectiveness of the model.

\subsection{Experimental Dataset}
Since we treat the recommendation task as a classification problem (Eq.\ref{eq:classification}), in order to train our model, we require positive and negative samples of items, where positive samples represent items relevant to the user and their shopping journey at impression time.  eBay's e-commerce site (and mobile apps) features millions of listing pages corresponding to active items, and each page contains recommendations for other items, organized into horizontal modules representing "similar items", "related items", "seller's other items", "items based on recent views" and so on. These recommendations are typically powered by module-specific recall and ranking stages. Each module presents multiple items (up to 12 on desktop web), and there may be as many as 6 such modules on each listing page, distributed along the length of the page.

In order to collect positive and negative data samples, we looked at implicit user interactions with these merchandising recommendation modules on eBay's listing pages, captured in the form of offline log data. Only those listing page impressions that had a recorded click event on a recommendation module were considered for positive and negative data samples. Recommended items across recommendations modules that were clicked on by a user were selected as positive examples for the model target. Click events were chosen due to the volume of available data, however, other signals such as purchases may also be used.  Recommended items that were not clicked on were treated as negative examples. Since clicking on a recommended item causes a new listing page to be loaded, each listing page impression typically resulted in one positive and multiple negative samples. As we shall discuss in the next section, the sampling strategy used for negative examples is critical to achieve good model training performance. 

The data needed for the user tower was gathered over a 30 day period going back from a given page impression. All of the positive and negative items were enriched with necessary metadata about category, title, and aspects using offline tables. A typical training run would consist of around 10 million page impressions gathered from 8 days of data. A validation set with approximately 110K page impressions was collected following the end of the training data time frame, in order to avoid information leakage across training and validation sets. In order to avoid biasing the outcome towards a few users with high engagement, a given user was only allowed to contribute to one page impression in the training data and validation data. Therefore, we had 10 million unique users and 110K unique users in our training and validation datasets respectively. In order to better capture the distribution of users and their diverse shopping patterns, data was collected from logs from all of eBay's platform experiences: desktop web, mobile web, and iOS and Android native apps.

\subsection{Negative Sampling} \label{sec:negative}
As previously mentioned, the number of available items $|V|$ on the eBay marketplace is on the scale of one billion, therefore it is infeasible to perform a full-size softmax operation as defined in Eq.\ref{eq:classification}. We experimented with two approaches for sampling negative examples.

\subsubsection{Observed Un-Clicked Item} In this approach, on the listing page, we take the item(s) clicked on as positive, and a subset of the items that were impressed but not clicked on as our negatives. Specifically, each positive item is paired with 8 un-clicked negative items. This approach failed in our initial model training, resulting in overfitted models that were unable to generalize. The main reason for this is that on the listing page, all of the impressed items from current recommendation modules are very similar to the seed item, and this leads to the effect that the model is unable distinguish positive from negative examples utilizing content-based item features.

\subsubsection{In-batch Random Negative Sampling} We then experimented with using random items as negatives. Rather than randomly sampling items from the whole item pool (billions of items), we use in-batch negative sampling \citep{hidasi2015session} by using the impressed but un-clicked items from other training examples within the same batch as negatives. This approach gives us a less complex and more efficient sampling strategy. This approach has some similarities with a popularity-based sampling approach, as the likelihood of an item serving as a  negative sample is proportional to the number of times this item is presented to a user. 

\subsection{Evaluation Metrics}
\label{sec:metrics}
We experimented with multiple evaluation metrics to measure model performance using our offline validation dataset. Given the similarity of our problem to the ranking problem in the information retrieval setting, we considered several metrics commonly used for ranking problems such as Normalized Discounted Cumulative Gain (NDCG), Recall@k, Precision@k, and Mean Reciprocal Rank (MRR)~\cite{freno:amazon}. As mentioned in the previous section, we typically have only 1 positive in each page impression, therefore it becomes important to measure whether or not the positive recommendation is in the top k results. We therefore ultimately chose to use Recall@k as our primary evaluation metric, for $k=1,5,10,20,40$. For P impressions, the metric is defined as:
\begin{equation}
    Recall@k = \frac{1}{P}\sum_{i=1}^{P}\frac{\textrm{\# relevant items @ k}}{\textrm{\# total relevant items}}
\end{equation}

For an industrial recommender system, it is important to surface the most relevant recommendations at the very top of the ranking since the user may only be shown (say) 5 recommendations via the user interface and would not engage with other recommendations.

\subsection{Model Training}

We used the PyTorch~\citep{NEURIPS2019_9015} deep learning framework to implement the core model. Additionally, we utilized the PyTorch-Lightning~\citep{falcon2019pytorch} framework which shortened development iterations and standardized the training loop so that it was seamless to transition the model between different CPU and GPU training environments. We chose the Adam optimizer~\citep{kingma2014adam} with a 0.01 learning rate. The gradient clipping parameter, set to 0.001, was essential in stabilizing the gradient in the recurrent part of the network, which spanned several hundred steps. We chose to sample 3000 negatives for each positive item, and use 600 as our batch size to maximize GPU utilization. Finally, we trained the model with 10 epochs over our data to reach convergence of the evaluation metrics. Model hyperparameters were selected considering production storage constraints and model performance on the validation set.  The chosen settings are reported in Table~\ref{table:hyperparameter}.

\begin{table}[t]
\begin{tabular}{@{}lll@{}}
\toprule
\textbf{Symbol}   & \textbf{Hyperparameter Description}      & \textbf{Value} \\ \midrule
$D$               & Item/User embedding dimension          & 64             \\
$D_{text}$        & Text-based feature dimension             & 64             \\
$D_{category}$    & Category feature dimenstion              & 64             \\
$L$               & Number of hidden layers in MLP            & 3              \\
$H$               & Hidden dimension in MLP        & 64             \\ 
$\tau$            & Temperature in affinity function         & 0.1          \\\bottomrule
\end{tabular}
\caption{Model hyperparameter settings.} \label{table:hyperparameter}
\end{table}

\subsection{Offline Evaluation}

As an offline baseline recommendation method for comparison, we used Recently Viewed Items (RVI), which recommends items that a user has recently viewed ranked by the viewed item’s recency. Although this method is simple and does not use a collaborative filtering (CF) based approach, RVI is widely used as a way of generating personalized recommendations in production systems. It is also a difficult baseline to beat in terms of generating user engagement, given that these are items the user has engaged with recently. The works of \citet{song2016multi} and \citet{wang2019attention} show approaches similar to RVI to be strong baseline methods, outperforming CF-based methods. 

\begin{table}[]
\begin{tabular}{@{}c|cc@{}}
\toprule
\multicolumn{1}{l|}{\textbf{Recall@k}} & \multicolumn{1}{l}{\textbf{RVI}} & \multicolumn{1}{l}{\textbf{Proposed Model}} \\ \midrule
1                             & 0.01                    & \textbf{0.02 (50\%)}                               \\
5                             & 0.06                    & 0.06                               \\
10                            & 0.09                    & 0.09                               \\
20                            & 0.12                    & \textbf{0.13 (8.3\%)}                               \\
40                            & 0.16                    & \textbf{0.18 (12.5\%)}                               \\ \bottomrule
\end{tabular} 
\caption{Offline test set evaluation results.} \label{table:offline_metric}
\end{table}

We evaluated our best model, which used a recurrent user representation based on item views and search query events, and the hyperparameters shown in Table~\ref{table:hyperparameter}. This evaluation was performed on a separate test set, which consists of 7K unique users and 10 million candidate items. The number of candidate items used for this evaluation is similar in scale to the number of candidate items typically used at prediction time in production. 

Experimental results are reported in Table~\ref{table:offline_metric}. Our method outperforms the RVI method in several Recall@k metrics that were measured. This shows that our model is able to generate appropriate personalized recommendations based on the user's current shopping mission, and potentially inspire new ones given the right user history. The RVI method, in contrast, only serves a re-targeting purpose, wherein a user is shown items they have already browsed in order to encourage re-engagement with previous shopping missions. In our approach, the multi-modal user embedding framework allows the model to incorporate various user activities seamlessly in a machine-learned manner to maximize user engagement. 

\subsection{Data Ablation Analysis and Model Robustness}

Stability of the model and robustness of its prediction is essential for a production environment, but is rarely studied for machine learning models that power recommender systems. We conducted an ablation analysis for our model with respect to user history data in order to study model performance under the condition where part of the user history is missing. 

\begin{figure}[h]
  \centering
   \includegraphics[width=\columnwidth]{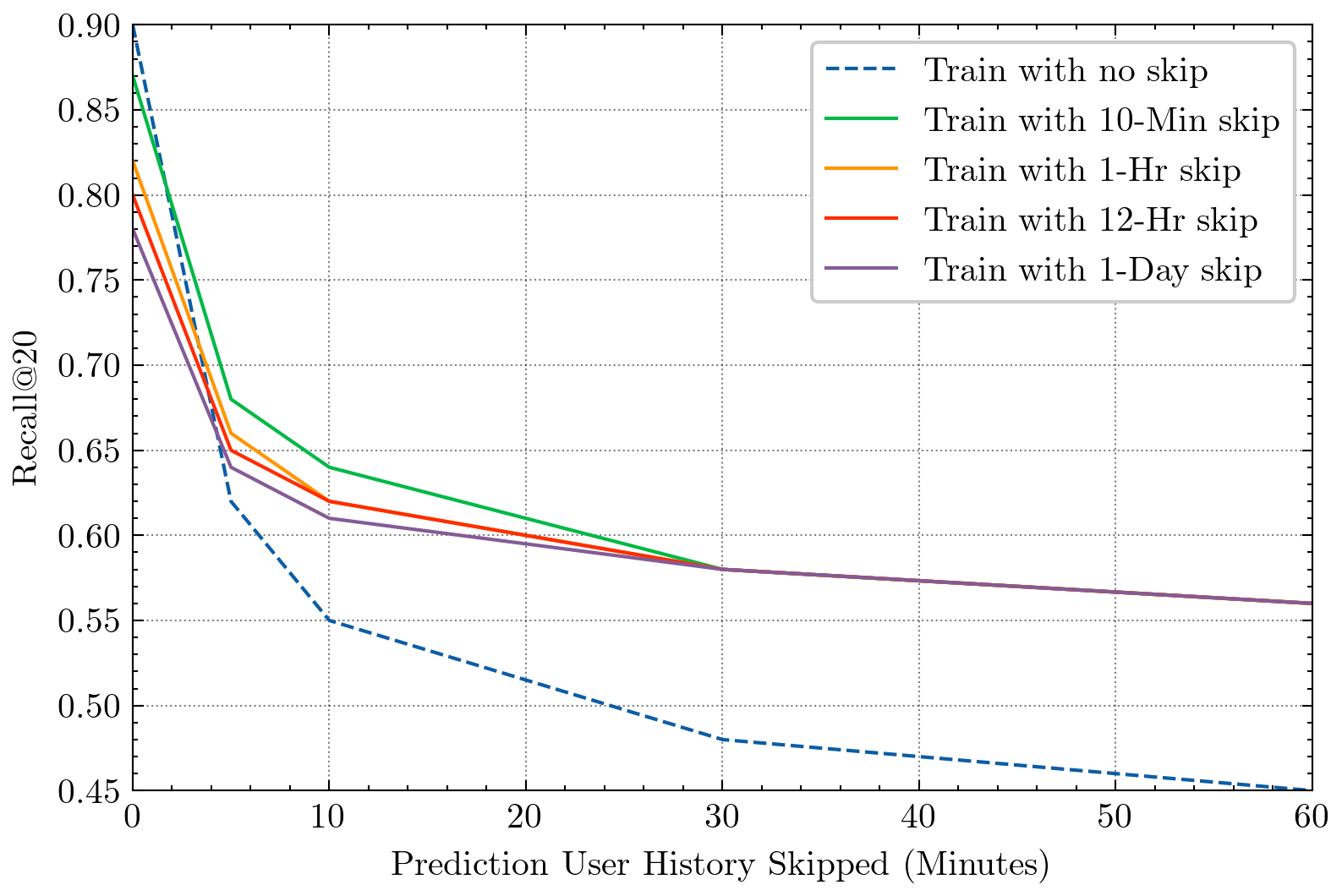}
  \caption{Model prediction performance with missing user history.}\label{fig:data_skip}
\end{figure}

To understand the possible impact of missing the most recent user history at prediction time, we performed several experiments, the results of which are shown in Figure~\ref{fig:data_skip}. First, we trained a model with the full user history present (dashed blue curve in Fig.~\ref{fig:data_skip}), and computed predictions on a validation set while dropping different lengths of the most recent user activity (horizontal axis in Fig.~\ref{fig:data_skip}). As can be seen by the dashed blue curve in Fig. \ref{fig:data_skip}, when the most recent 5-minute user activity was missing, the Recall@20 metric decreased by more than 30\%, from 0.9 to 0.62. The metric degrades by as much as 50\% to 0.45 when user activity within the most recent 60 minutes is missing. This creates significant performance risk for production deployment, since the model may not always have up-to-date real-time user onsite history at prediction time. To counteract this effect, we chose to train the model while dropping the most recent user activities, not random ones from the user history, in order to better align with the scenario happening in production system where there may be a gap in time between the batch model prediction output and a user impression. A user can simply be browsing the site, potentially with a new shopping mission, for some time after a batch update.

We experimented with training models by dropping some of the user onsite activity data before impression time, ranging from the most recent 10 minutes (green curve) to 1 day (purple curve). As we can see in Fig.~\ref{fig:data_skip}, models trained using skipped user history performed better compared to the original model, when part of user history was missing at prediction time. For example, with 60 minutes of user activity missing at prediction time, all "skipped" models were able to achieve Recall@20 of 0.56, compared to 0.45 for the original model (dashed blue curve). In addition, training with more skipped history leads to a "flatter" curve, suggesting a more robust model under conditions of variable missing history. However, we also observe that model performance decreases when trained with more dropped out user history, especially across the range of 0 minute (no skipping) to 30 minutes. In order to find a balance between performance and consistency, the production model is selected as the one with the largest area under curve from Fig. \ref{fig:data_skip} amongst those trained with 10-minutes of skipped user activity (green curve). This sort of user history dropout is important to consider when training a model with robust prediction expectations for a production setting.

\section{Model Prediction}
\label{sec:eng}
The previous sections focused on describing the model and offline validation testing. In this section, we will turn our attention to describing the model prediction stage, including the multitude of engineering considerations and trade-offs for building a large scale production engineering recommender system.

\subsection{Retrieval}
\label{sec:knn}
During the prediction stage, given a user embedding and a pool of candidate item embeddings, retrieval is conducted by the KNN search algorithm. We used the KNN implementation from FAISS~\citep{JDH17}. For a marketplace with an enormous item-based inventory, similar item listings are common on the site (at the time of paper writing, searching ``iphone 11'' on eBay would return 3047 results). As our model only consumes content-based features (title, category, and aspects) for items, all those content-similar items would have similar embedding from our model. Utilizing the traditional KNN approach, given a user embedding, the retrieved items would be extremely overlapping in the embedding space. 

\begin{figure}[h]
  \includegraphics[width=\linewidth]{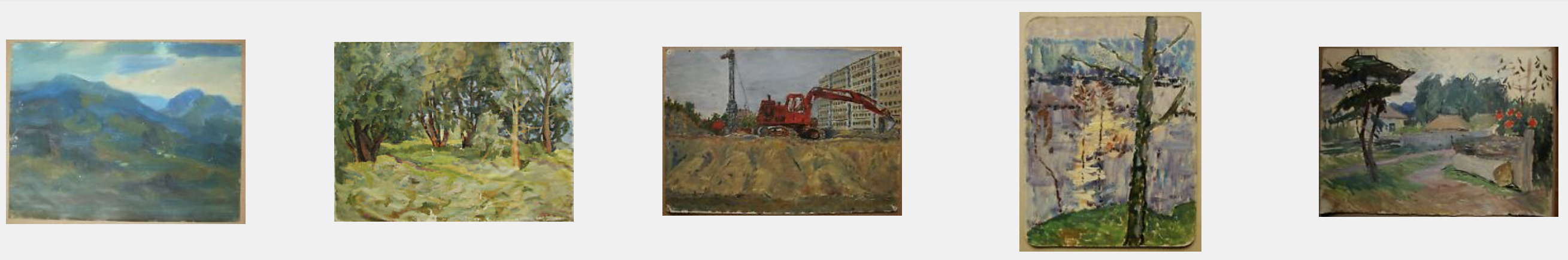}
  \includegraphics[width=\linewidth]{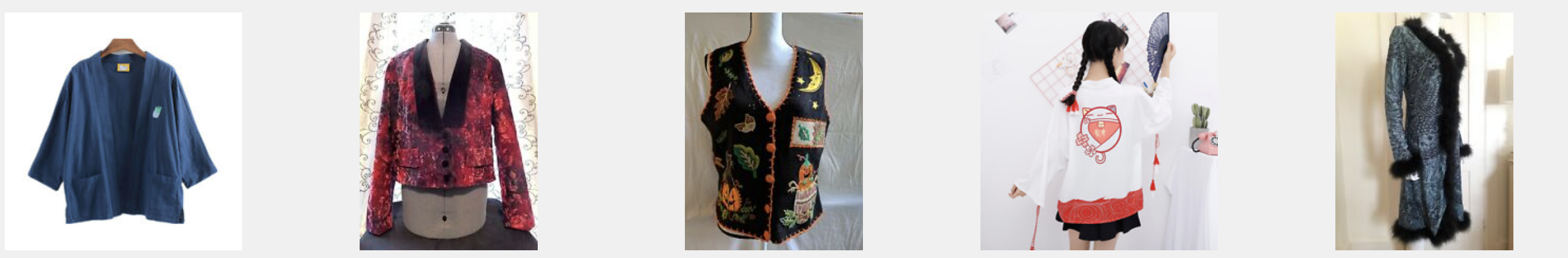}
  \includegraphics[width=\linewidth]{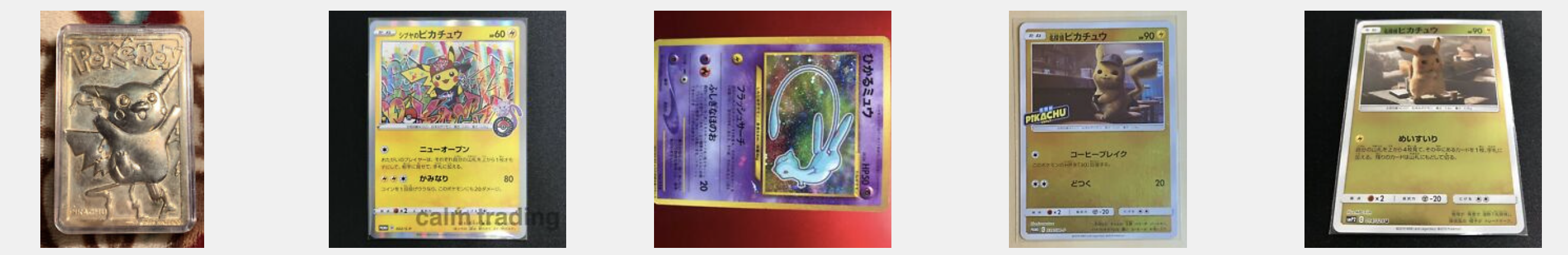}
  \caption{Sample clusters demonstrating model quality.}
  \label{fig:cluster}
\end{figure}

To address this diversity problem, we use K-means clustering to group all candidate items into $K$ clusters, each with a centroid $c_i$. At retrieval time, we try to find $N$ candidate recall items, given a user embedding $u$. We first find the nearest $M$ clusters, and in each cluster conduct a KNN search to retrieve $m_i$ items:
\begin{equation}
    m_i = \lceil \frac{e^{\gamma(\mathbf{c}_i, \mathbf{u})}}{\sum_{j=1}^{M}e^{\gamma(\mathbf{c}_j, \mathbf{u})}} \cdot N \rceil,
\end{equation}
where $\gamma(\mathbf{c_j}, \mathbf{u})$ is the same affinity function defined in Eq. \ref{eq:classification}. Figure~\ref{fig:cluster} demonstrates the quality of the item clusters generated by the model, with each row representing one cluster.

\begin{figure}[h]
  \includegraphics[width=\linewidth]{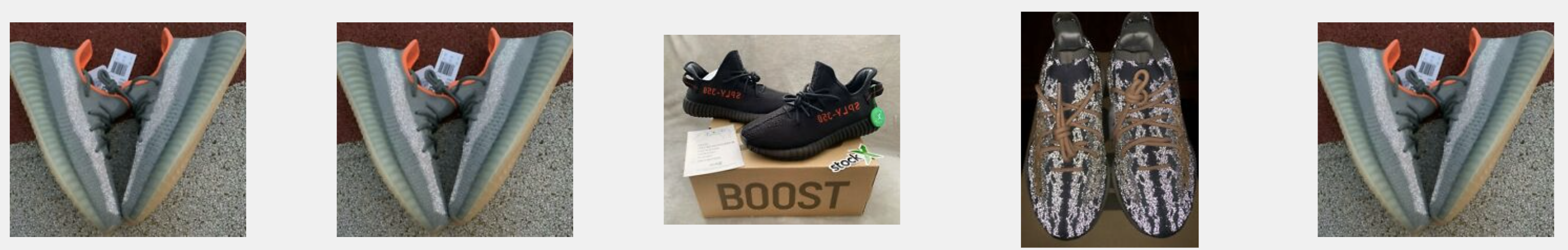}
  \centering
  \small
  (a) Without Clustering ($M=1$, $K$=100,000)
  \includegraphics[width=\linewidth]{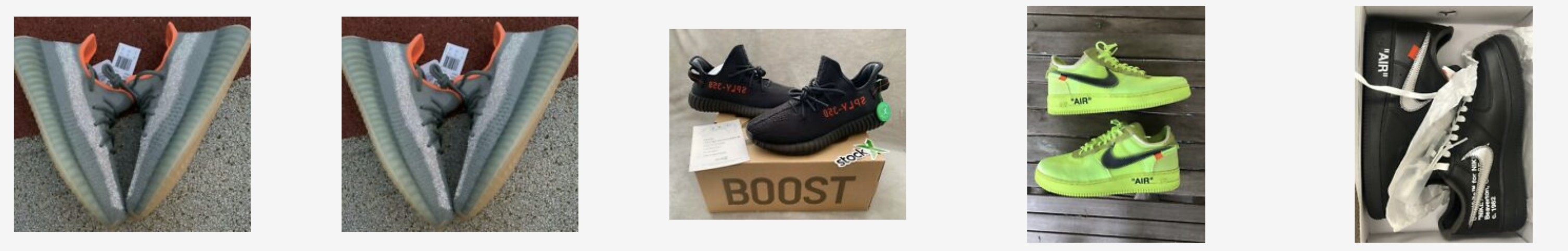}
  \centering
  \small
  (b) With Clustering ($M=10$, $K$=100,000)
  \caption{Generated personalized item recommendations (a) without and (b) with clustering, where the user has been previously looking at Adidas Yeezy shoes.}
  \label{fig:cluster_before_after}
\end{figure}

Since the inventory on eBay is highly diverse, the existing product catalog does not cover all eBay items consistently. The clustering step essentially creates a "pseudo catalog" covering all items, and content-similar items could be organized into static entities. In our experiment, we found $K$ = 100,000 provides the right level of granularity in clustering items. 

The number of clusters, $M$, for retrieval is chosen by balancing between item diversity and retrieval metrics. With larger $M$, the retrieved items would cover more potential user interests, but with less concentration on a specific direction. With $M = 1$, this approach degenerates to the traditional KNN method. In our experiment, as can been shown from an example in Fig. \ref{fig:cluster_before_after}, the clustering-based method generates a more diverse set of item recommendations, without losing relevancy. This technique enables controlling recommendation diversity with a simple hyperparameter, $M$, and can be tuned during the prediction stage separately from the training stage. In production, we found that using $M = 10$ and $K$ = 100,000 provides the optimal amount of diversity of eBay item recommendations. 

\subsection{Production Engineering Architecture}
\begin{figure}
  \centering
  \includegraphics[width=\linewidth]{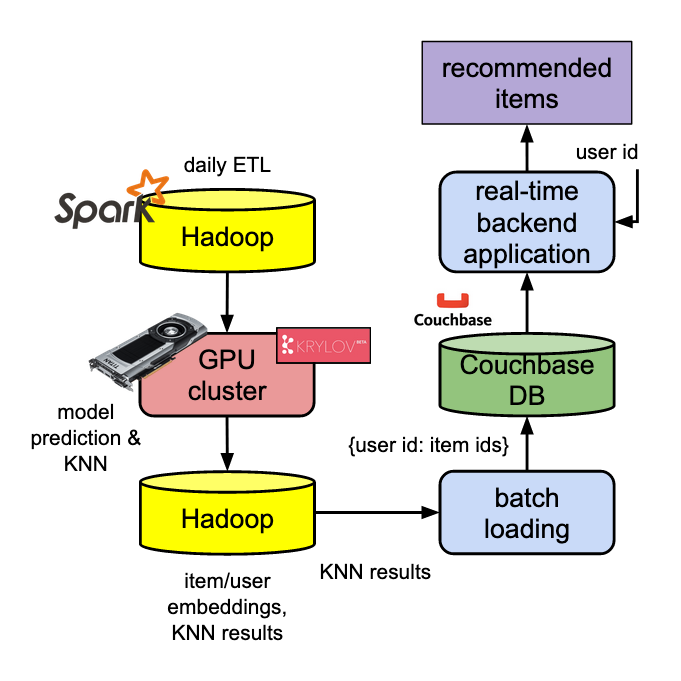}
  \caption{Production engineering architecture for model prediction.}
  \label{fig:eng}
\end{figure}

In this section, we describe the engineering architecture used for model prediction to serve the personalized recommended items to eBay users. The production engineering architecture is depicted in Figure~\ref{fig:eng}. Since a user's browsing history is constantly being updated, we recalculate the user embedding as well as the KNN results on a daily basis. Note that our model is based on content text based features, which are mostly static for any given item, so we found that we do not need to retrain the full model on a daily basis.

The prediction process is performed offline in batch mode. First, two Spark extract, transform, load (ETL) jobs generate the candidate items and the up to date user histories for all users on eBay that had activity in the last 30 days, along with the necessary metadata aggregated in Hadoop. The user behaviour data at eBay is aggregated once per day in Hadoop, so this is the reason the ETL jobs run with this daily frequency. In order to control for item popularity, we limit the candidate item set to items that have had 2 or more clicks in the past 4 days.

Next, we utilize eBay's GPU cluster, Krylov~\cite{krylov1}, to run a forward pass on the item and user towers in the trained model, in order to generate the item and user embeddings, respectively. Now, for every user embedding, a KNN search is performed on all of the item embeddings, to generate the KNN results and write them back to Hadoop. The KNN results are then loaded to a Couchbase database, which is utilized as a fast (with a latency of a few milliseconds) run-time cache, with the user id as the key, using a batch loading application. This caching approach is scalable to hundreds of millions of users, and is only limited by the Couchbase cluster capacity being used.

The eBay run-time web serving application stack is based on the Java Virtual Machine (JVM). The backend application serving recommendations is written in Scala and runs on the JVM for fast run-time performance. One of the reasons the caching architecture was chosen, is due to the latency requirements of the run-time application for serving personalized recommendations to the eBay users. Traditional approaches, such as in-memory matrix factorization, would simply not be computationally feasible at eBay data scale. In addition to enriching the recommended items with necessary metadata, the backend application can also apply a separately trained LTR model~\cite{Brovman2} in order to optimize conversion.

\subsection{Online Evaluation}
The best trained model was deployed to our production environment and evaluated in an online A/B test. Compared to the current personalized recommendation module in production, which mainly focuses on resurfacing items a user has viewed, our model demonstrated an increase in the surface rate of $\sim$6\% to generate recommendations for 90\% of listing page impressions. This is mainly achieved by addressing the cold start and data sparsity problem through content-based item and user embeddings. The production RVI based algorithm is limited to using a user's item activity only, as opposed to our model which uses search activity as well. Additionally, for the production algorithm, specific items can be filtered out due to item expiration or business logic. The embeddings based approach generates a much larger candidate set, which results in a higher surface rate overall. The current system, refreshed daily to capture new user activity, serves 50 million US users and will be scaled to cover all global eBay users in the near future.

\section{Summary and Future Work}
\label{sec:conc}
In this paper, we presented an approach for generating personalized item recommendations in a large scale e-commerce marketplace. A two-tower deep learning model is used to learn embeddings of items and users in a shared vector space. Items and users embeddings are learned using their content features and multi-modal onsite user activity respectively, to tackle the cold-start problem. To better address the instability of the online production environment, user data ablation is incorporated into the offline training process to generate a more robust model. Offline and online experiments have validated our approach, and shown significant improvements over baseline approaches. A personalized recommender system based on our approach has been launched to production and is now serving recommendations at scale to eBay buyers. 

We are actively working to enhance the quality of the model. More types of implicit user feedback, such as "add to cart" and "watch item" should be incorporated into the model to better capture a user's shopping mission. Additionally, having only a single vector to represent a user is potentially limiting, since the user may have diverse interests and multiple ongoing shopping missions. We are working on an improved user representation that encodes parallel shopping missions, and has the capability to separate long term user interests from short term shopping missions. From the item modeling perspective, a pre-trained language understanding model (e.g BERT \citep{devlin2018bert}) could be leveraged to better understand item titles. We are also working on incorporating an additional LTR model at runtime which to improve conversion metrics. This LTR model would be trained specifically for our context wherein the input is the user history and not a $seed$ item.   

One of the shortcomings of our current engineering implementation is the frequency of the recommendation update, which is currently limited by the need for daily batch processing. We are working on moving to a near real-time (NRT) system for generating personalized recommendations using user and item embeddings. This entails several infrastructural improvements including i) a real-time KNN service based on algorithms such as FAISS~\cite{JDH17} and ScaNN~\cite{avq_2020} to compute distances between embedding vectors, ii) a real-time model prediction service for generating item and user embeddings (potentially using the Open Neural Network Exchange (ONNX) format~\cite{bai2019} for transferring the trained model from the offline training to the real-time prediction environment), and iii) an event stream processing service for capturing up to date item and user actions on the eBay site. We are actively working on developing this infrastructure in the form of more general services that can be utilized for a variety of deep learning embedding-based algorithms, in addition to personalized recommendations.

\begin{acks}
We would like to thank Jesse Lute for product guidance and A/B testing. We also appreciate useful feedback on the manuscript from Zhe Wu, Hongliang Yu, and Menghan Wang.
\end{acks}

\bibliographystyle{ACM-Reference-Format}
\bibliography{bibliography}


\begin{thebibliography}{33}


\ifx \showCODEN    \undefined \def \showCODEN     #1{\unskip}     \fi
\ifx \showDOI      \undefined \def \showDOI       #1{#1}\fi
\ifx \showISBNx    \undefined \def \showISBNx     #1{\unskip}     \fi
\ifx \showISBNxiii \undefined \def \showISBNxiii  #1{\unskip}     \fi
\ifx \showISSN     \undefined \def \showISSN      #1{\unskip}     \fi
\ifx \showLCCN     \undefined \def \showLCCN      #1{\unskip}     \fi
\ifx \shownote     \undefined \def \shownote      #1{#1}          \fi
\ifx \showarticletitle \undefined \def \showarticletitle #1{#1}   \fi
\ifx \showURL      \undefined \def \showURL       {\relax}        \fi
\providecommand\bibfield[2]{#2}
\providecommand\bibinfo[2]{#2}
\providecommand\natexlab[1]{#1}
\providecommand\showeprint[2][]{arXiv:#2}

\bibitem[\protect\citeauthoryear{Aggarwal}{Aggarwal}{2016}]%
        {recsys1}
\bibfield{author}{\bibinfo{person}{Charu~C. Aggarwal}.}
  \bibinfo{year}{2016}\natexlab{}.
\newblock \bibinfo{booktitle}{\emph{Recommender Systems: The Textbook}
  (\bibinfo{edition}{1st} ed.)}.
\newblock \bibinfo{publisher}{Springer Publishing Company, Incorporated}.
\newblock
\showISBNx{3319296574}


\bibitem[\protect\citeauthoryear{Bai, Lu, Zhang, et~al\mbox{.}}{Bai
  et~al\mbox{.}}{2019}]%
        {bai2019}
\bibfield{author}{\bibinfo{person}{Junjie Bai}, \bibinfo{person}{Fang Lu},
  \bibinfo{person}{Ke Zhang}, {et~al\mbox{.}}} \bibinfo{year}{2019}\natexlab{}.
\newblock \bibinfo{title}{ONNX: Open Neural Network Exchange}.
\newblock \bibinfo{howpublished}{\url{https://github.com/onnx/onnx}}.
\newblock


\bibitem[\protect\citeauthoryear{Brovman}{Brovman}{2019}]%
        {Brovman1}
\bibfield{author}{\bibinfo{person}{Y.~M. Brovman}.}
  \bibinfo{year}{2019}\natexlab{}.
\newblock \bibinfo{booktitle}{\emph{Complementary Item Recommendations at eBay
  Scale}}.
\newblock
\urldef\tempurl%
\url{https://tech.ebayinc.com/engineering/complementary-item-recommendations-at-ebay-scale/}
\showURL{%
\tempurl}


\bibitem[\protect\citeauthoryear{Brovman, Jacob, Srinivasan, Neola, Galron,
  Snyder, and Wang}{Brovman et~al\mbox{.}}{2016}]%
        {Brovman2}
\bibfield{author}{\bibinfo{person}{Yuri~M. Brovman}, \bibinfo{person}{Marie
  Jacob}, \bibinfo{person}{Natraj Srinivasan}, \bibinfo{person}{Stephen Neola},
  \bibinfo{person}{Daniel Galron}, \bibinfo{person}{Ryan Snyder}, {and}
  \bibinfo{person}{Paul Wang}.} \bibinfo{year}{2016}\natexlab{}.
\newblock \showarticletitle{Optimizing Similar Item Recommendations in a
  Semi-Structured Marketplace to Maximize Conversion}. In
  \bibinfo{booktitle}{\emph{Proceedings of the 10th ACM Conference on
  Recommender Systems}} (Boston, Massachusetts, USA)
  \emph{(\bibinfo{series}{RecSys '16})}. \bibinfo{publisher}{Association for
  Computing Machinery}, \bibinfo{address}{New York, NY, USA},
  \bibinfo{pages}{199–202}.
\newblock
\showISBNx{9781450340359}
\urldef\tempurl%
\url{https://doi.org/10.1145/2959100.2959166}
\showDOI{\tempurl}


\bibitem[\protect\citeauthoryear{Chen, Zhang, Lu, Chen, Zheng, and Yu}{Chen
  et~al\mbox{.}}{2012}]%
        {chen2012svdfeature}
\bibfield{author}{\bibinfo{person}{Tianqi Chen}, \bibinfo{person}{Weinan
  Zhang}, \bibinfo{person}{Qiuxia Lu}, \bibinfo{person}{Kailong Chen},
  \bibinfo{person}{Zhao Zheng}, {and} \bibinfo{person}{Yong Yu}.}
  \bibinfo{year}{2012}\natexlab{}.
\newblock \showarticletitle{SVDFeature: a toolkit for feature-based
  collaborative filtering}.
\newblock \bibinfo{journal}{\emph{The Journal of Machine Learning Research}}
  \bibinfo{volume}{13}, \bibinfo{number}{1} (\bibinfo{year}{2012}),
  \bibinfo{pages}{3619--3622}.
\newblock


\bibitem[\protect\citeauthoryear{Cheng, Koc, Harmsen, Shaked, Chandra, Aradhye,
  Anderson, Corrado, Chai, Ispir, et~al\mbox{.}}{Cheng et~al\mbox{.}}{2016}]%
        {cheng2016wide}
\bibfield{author}{\bibinfo{person}{Heng-Tze Cheng}, \bibinfo{person}{Levent
  Koc}, \bibinfo{person}{Jeremiah Harmsen}, \bibinfo{person}{Tal Shaked},
  \bibinfo{person}{Tushar Chandra}, \bibinfo{person}{Hrishi Aradhye},
  \bibinfo{person}{Glen Anderson}, \bibinfo{person}{Greg Corrado},
  \bibinfo{person}{Wei Chai}, \bibinfo{person}{Mustafa Ispir}, {et~al\mbox{.}}}
  \bibinfo{year}{2016}\natexlab{}.
\newblock \showarticletitle{Wide \& deep learning for recommender systems}. In
  \bibinfo{booktitle}{\emph{Proceedings of the 1st workshop on deep learning
  for recommender systems}}. \bibinfo{pages}{7--10}.
\newblock


\bibitem[\protect\citeauthoryear{Cho, Van~Merri{\"e}nboer, Gulcehre, Bahdanau,
  Bougares, Schwenk, and Bengio}{Cho et~al\mbox{.}}{2014}]%
        {cho2014learning}
\bibfield{author}{\bibinfo{person}{Kyunghyun Cho}, \bibinfo{person}{Bart
  Van~Merri{\"e}nboer}, \bibinfo{person}{Caglar Gulcehre},
  \bibinfo{person}{Dzmitry Bahdanau}, \bibinfo{person}{Fethi Bougares},
  \bibinfo{person}{Holger Schwenk}, {and} \bibinfo{person}{Yoshua Bengio}.}
  \bibinfo{year}{2014}\natexlab{}.
\newblock \showarticletitle{Learning phrase representations using RNN
  encoder-decoder for statistical machine translation}.
\newblock \bibinfo{journal}{\emph{arXiv preprint arXiv:1406.1078}}
  (\bibinfo{year}{2014}).
\newblock


\bibitem[\protect\citeauthoryear{Covington, Adams, and Sargin}{Covington
  et~al\mbox{.}}{2016}]%
        {covington2016deep}
\bibfield{author}{\bibinfo{person}{Paul Covington}, \bibinfo{person}{Jay
  Adams}, {and} \bibinfo{person}{Emre Sargin}.}
  \bibinfo{year}{2016}\natexlab{}.
\newblock \showarticletitle{Deep neural networks for youtube recommendations}.
  In \bibinfo{booktitle}{\emph{Proceedings of the 10th ACM conference on
  recommender systems}}. \bibinfo{pages}{191--198}.
\newblock


\bibitem[\protect\citeauthoryear{Devlin, Chang, Lee, and Toutanova}{Devlin
  et~al\mbox{.}}{2018}]%
        {devlin2018bert}
\bibfield{author}{\bibinfo{person}{Jacob Devlin}, \bibinfo{person}{Ming-Wei
  Chang}, \bibinfo{person}{Kenton Lee}, {and} \bibinfo{person}{Kristina
  Toutanova}.} \bibinfo{year}{2018}\natexlab{}.
\newblock \showarticletitle{Bert: Pre-training of deep bidirectional
  transformers for language understanding}.
\newblock \bibinfo{journal}{\emph{arXiv preprint arXiv:1810.04805}}
  (\bibinfo{year}{2018}).
\newblock


\bibitem[\protect\citeauthoryear{Falcon}{Falcon}{2019}]%
        {falcon2019pytorch}
\bibfield{author}{\bibinfo{person}{WA Falcon}.}
  \bibinfo{year}{2019}\natexlab{}.
\newblock \showarticletitle{PyTorch Lightning}.
\newblock \bibinfo{journal}{\emph{GitHub. Note:
  https://github.com/PyTorchLightning/pytorch-lightning}}  \bibinfo{volume}{3}
  (\bibinfo{year}{2019}).
\newblock


\bibitem[\protect\citeauthoryear{Freno et~al\mbox{.}}{Freno
  et~al\mbox{.}}{2015}]%
        {freno:amazon}
\bibfield{author}{\bibinfo{person}{Antonino Freno} {et~al\mbox{.}}}
  \bibinfo{year}{2015}\natexlab{}.
\newblock \showarticletitle{One-Pass Ranking Models for Low-Latency Product
  Recommendations}. In \bibinfo{booktitle}{\emph{SIGKDD}} (Sydney, NSW,
  Australia). 10.
\newblock
\showISBNx{978-1-4503-3664-2}
\urldef\tempurl%
\url{https://doi.org/10.1145/2783258.2788579}
\showDOI{\tempurl}


\bibitem[\protect\citeauthoryear{Galron, Brovman, Chung, Wieja, and
  Wang}{Galron et~al\mbox{.}}{2018}]%
        {galron2018deep}
\bibfield{author}{\bibinfo{person}{Daniel~A Galron}, \bibinfo{person}{Yuri~M
  Brovman}, \bibinfo{person}{Jin Chung}, \bibinfo{person}{Michal Wieja}, {and}
  \bibinfo{person}{Paul Wang}.} \bibinfo{year}{2018}\natexlab{}.
\newblock \showarticletitle{Deep Item-based Collaborative Filtering for Sparse
  Implicit Feedback}.
\newblock \bibinfo{journal}{\emph{arXiv preprint arXiv:1812.10546}}
  (\bibinfo{year}{2018}).
\newblock


\bibitem[\protect\citeauthoryear{Gao, Fan, Sun, Jia, Xiao, Wang, and Liu}{Gao
  et~al\mbox{.}}{2020}]%
        {gao2020deep}
\bibfield{author}{\bibinfo{person}{Weihao Gao}, \bibinfo{person}{Xiangjun Fan},
  \bibinfo{person}{Jiankai Sun}, \bibinfo{person}{Kai Jia},
  \bibinfo{person}{Wenzhi Xiao}, \bibinfo{person}{Chong Wang}, {and}
  \bibinfo{person}{Xiaobing Liu}.} \bibinfo{year}{2020}\natexlab{}.
\newblock \showarticletitle{Deep Retrieval: An End-to-End Learnable Structure
  Model for Large-Scale Recommendations}.
\newblock \bibinfo{journal}{\emph{arXiv preprint arXiv:2007.07203}}
  (\bibinfo{year}{2020}).
\newblock


\bibitem[\protect\citeauthoryear{Glorot, Bordes, and Bengio}{Glorot
  et~al\mbox{.}}{2011}]%
        {glorot2011deep}
\bibfield{author}{\bibinfo{person}{Xavier Glorot}, \bibinfo{person}{Antoine
  Bordes}, {and} \bibinfo{person}{Yoshua Bengio}.}
  \bibinfo{year}{2011}\natexlab{}.
\newblock \showarticletitle{Deep sparse rectifier neural networks}. In
  \bibinfo{booktitle}{\emph{Proceedings of the fourteenth international
  conference on artificial intelligence and statistics}}.
  \bibinfo{pages}{315--323}.
\newblock


\bibitem[\protect\citeauthoryear{Guo, Sun, Lindgren, Geng, Simcha, Chern, and
  Kumar}{Guo et~al\mbox{.}}{2020}]%
        {avq_2020}
\bibfield{author}{\bibinfo{person}{Ruiqi Guo}, \bibinfo{person}{Philip Sun},
  \bibinfo{person}{Erik Lindgren}, \bibinfo{person}{Quan Geng},
  \bibinfo{person}{David Simcha}, \bibinfo{person}{Felix Chern}, {and}
  \bibinfo{person}{Sanjiv Kumar}.} \bibinfo{year}{2020}\natexlab{}.
\newblock \showarticletitle{Accelerating Large-Scale Inference with Anisotropic
  Vector Quantization}. In \bibinfo{booktitle}{\emph{International Conference
  on Machine Learning}}.
\newblock
\urldef\tempurl%
\url{https://arxiv.org/abs/1908.10396}
\showURL{%
\tempurl}


\bibitem[\protect\citeauthoryear{He, Liao, Zhang, Nie, Hu, and Chua}{He
  et~al\mbox{.}}{2017}]%
        {he2017neural}
\bibfield{author}{\bibinfo{person}{Xiangnan He}, \bibinfo{person}{Lizi Liao},
  \bibinfo{person}{Hanwang Zhang}, \bibinfo{person}{Liqiang Nie},
  \bibinfo{person}{Xia Hu}, {and} \bibinfo{person}{Tat-Seng Chua}.}
  \bibinfo{year}{2017}\natexlab{}.
\newblock \showarticletitle{Neural collaborative filtering}. In
  \bibinfo{booktitle}{\emph{Proceedings of the 26th international conference on
  world wide web}}. \bibinfo{pages}{173--182}.
\newblock


\bibitem[\protect\citeauthoryear{Hidasi, Karatzoglou, Baltrunas, and
  Tikk}{Hidasi et~al\mbox{.}}{2015}]%
        {hidasi2015session}
\bibfield{author}{\bibinfo{person}{Bal{\'a}zs Hidasi},
  \bibinfo{person}{Alexandros Karatzoglou}, \bibinfo{person}{Linas Baltrunas},
  {and} \bibinfo{person}{Domonkos Tikk}.} \bibinfo{year}{2015}\natexlab{}.
\newblock \showarticletitle{Session-based recommendations with recurrent neural
  networks}.
\newblock \bibinfo{journal}{\emph{arXiv preprint arXiv:1511.06939}}
  (\bibinfo{year}{2015}).
\newblock


\bibitem[\protect\citeauthoryear{Hu, Koren, and Volinsky}{Hu
  et~al\mbox{.}}{2008}]%
        {hu2008collaborative}
\bibfield{author}{\bibinfo{person}{Yifan Hu}, \bibinfo{person}{Yehuda Koren},
  {and} \bibinfo{person}{Chris Volinsky}.} \bibinfo{year}{2008}\natexlab{}.
\newblock \showarticletitle{Collaborative filtering for implicit feedback
  datasets}. In \bibinfo{booktitle}{\emph{Data Mining, 2008. ICDM'08. Eighth
  IEEE International Conference on}}. Ieee, \bibinfo{pages}{263--272}.
\newblock


\bibitem[\protect\citeauthoryear{Johnson, Douze, and J{\'e}gou}{Johnson
  et~al\mbox{.}}{2017}]%
        {JDH17}
\bibfield{author}{\bibinfo{person}{Jeff Johnson}, \bibinfo{person}{Matthijs
  Douze}, {and} \bibinfo{person}{Herv{\'e} J{\'e}gou}.}
  \bibinfo{year}{2017}\natexlab{}.
\newblock \showarticletitle{Billion-scale similarity search with GPUs}.
\newblock \bibinfo{journal}{\emph{arXiv preprint arXiv:1702.08734}}
  (\bibinfo{year}{2017}).
\newblock


\bibitem[\protect\citeauthoryear{Katariya and Ramani}{Katariya and
  Ramani}{2019}]%
        {krylov1}
\bibfield{author}{\bibinfo{person}{S. Katariya} {and} \bibinfo{person}{A.
  Ramani}.} \bibinfo{year}{2019}\natexlab{}.
\newblock \bibinfo{booktitle}{\emph{eBay’s Transformation to a Modern AI
  Platform}}.
\newblock
\urldef\tempurl%
\url{https://tech.ebayinc.com/engineering/ebays-transformation-to-a-modern-ai-platform/}
\showURL{%
\tempurl}


\bibitem[\protect\citeauthoryear{Kingma and Ba}{Kingma and Ba}{2014}]%
        {kingma2014adam}
\bibfield{author}{\bibinfo{person}{Diederik~P Kingma} {and}
  \bibinfo{person}{Jimmy Ba}.} \bibinfo{year}{2014}\natexlab{}.
\newblock \showarticletitle{Adam: A method for stochastic optimization}.
\newblock \bibinfo{journal}{\emph{arXiv preprint arXiv:1412.6980}}
  (\bibinfo{year}{2014}).
\newblock


\bibitem[\protect\citeauthoryear{Koren, Bell, and Volinsky}{Koren
  et~al\mbox{.}}{2009}]%
        {koren2009matrix}
\bibfield{author}{\bibinfo{person}{Yehuda Koren}, \bibinfo{person}{Robert
  Bell}, {and} \bibinfo{person}{Chris Volinsky}.}
  \bibinfo{year}{2009}\natexlab{}.
\newblock \showarticletitle{Matrix factorization techniques for recommender
  systems}.
\newblock \bibinfo{journal}{\emph{Computer}} \bibinfo{volume}{42},
  \bibinfo{number}{8} (\bibinfo{year}{2009}).
\newblock


\bibitem[\protect\citeauthoryear{Melville, Mooney, and Nagarajan}{Melville
  et~al\mbox{.}}{2002}]%
        {melville2002content}
\bibfield{author}{\bibinfo{person}{Prem Melville}, \bibinfo{person}{Raymond~J
  Mooney}, {and} \bibinfo{person}{Ramadass Nagarajan}.}
  \bibinfo{year}{2002}\natexlab{}.
\newblock \showarticletitle{Content-boosted collaborative filtering for
  improved recommendations}.
\newblock \bibinfo{journal}{\emph{Aaai/iaai}}  \bibinfo{volume}{23}
  (\bibinfo{year}{2002}), \bibinfo{pages}{187--192}.
\newblock


\bibitem[\protect\citeauthoryear{Mikolov, Chen, Corrado, and Dean}{Mikolov
  et~al\mbox{.}}{2013}]%
        {mikolov2013efficient}
\bibfield{author}{\bibinfo{person}{Tomas Mikolov}, \bibinfo{person}{Kai Chen},
  \bibinfo{person}{Greg Corrado}, {and} \bibinfo{person}{Jeffrey Dean}.}
  \bibinfo{year}{2013}\natexlab{}.
\newblock \showarticletitle{Efficient estimation of word representations in
  vector space}.
\newblock \bibinfo{journal}{\emph{arXiv preprint arXiv:1301.3781}}
  (\bibinfo{year}{2013}).
\newblock


\bibitem[\protect\citeauthoryear{Paszke, Gross, Massa, Lerer, Bradbury, Chanan,
  Killeen, Lin, Gimelshein, Antiga, Desmaison, Kopf, Yang, DeVito, Raison,
  Tejani, Chilamkurthy, Steiner, Fang, Bai, and Chintala}{Paszke
  et~al\mbox{.}}{2019}]%
        {NEURIPS2019_9015}
\bibfield{author}{\bibinfo{person}{Adam Paszke}, \bibinfo{person}{Sam Gross},
  \bibinfo{person}{Francisco Massa}, \bibinfo{person}{Adam Lerer},
  \bibinfo{person}{James Bradbury}, \bibinfo{person}{Gregory Chanan},
  \bibinfo{person}{Trevor Killeen}, \bibinfo{person}{Zeming Lin},
  \bibinfo{person}{Natalia Gimelshein}, \bibinfo{person}{Luca Antiga},
  \bibinfo{person}{Alban Desmaison}, \bibinfo{person}{Andreas Kopf},
  \bibinfo{person}{Edward Yang}, \bibinfo{person}{Zachary DeVito},
  \bibinfo{person}{Martin Raison}, \bibinfo{person}{Alykhan Tejani},
  \bibinfo{person}{Sasank Chilamkurthy}, \bibinfo{person}{Benoit Steiner},
  \bibinfo{person}{Lu Fang}, \bibinfo{person}{Junjie Bai}, {and}
  \bibinfo{person}{Soumith Chintala}.} \bibinfo{year}{2019}\natexlab{}.
\newblock \showarticletitle{PyTorch: An Imperative Style, High-Performance Deep
  Learning Library}.
\newblock In \bibinfo{booktitle}{\emph{Advances in Neural Information
  Processing Systems 32}}, \bibfield{editor}{\bibinfo{person}{H.~Wallach},
  \bibinfo{person}{H.~Larochelle}, \bibinfo{person}{A.~Beygelzimer},
  \bibinfo{person}{F.~d\textquotesingle Alch\'{e}-Buc},
  \bibinfo{person}{E.~Fox}, {and} \bibinfo{person}{R.~Garnett}} (Eds.).
  \bibinfo{publisher}{Curran Associates, Inc.}, \bibinfo{pages}{8024--8035}.
\newblock
\urldef\tempurl%
\url{http://papers.neurips.cc/paper/9015-pytorch-an-imperative-style-high-performance-deep-learning-library.pdf}
\showURL{%
\tempurl}


\bibitem[\protect\citeauthoryear{Rendle}{Rendle}{2010}]%
        {rendle2010factorization}
\bibfield{author}{\bibinfo{person}{Steffen Rendle}.}
  \bibinfo{year}{2010}\natexlab{}.
\newblock \showarticletitle{Factorization machines}. In
  \bibinfo{booktitle}{\emph{2010 IEEE International Conference on Data
  Mining}}. IEEE, \bibinfo{pages}{995--1000}.
\newblock


\bibitem[\protect\citeauthoryear{Rendle, Freudenthaler, Gantner, and
  Schmidt-Thieme}{Rendle et~al\mbox{.}}{2009}]%
        {rendle2009bpr}
\bibfield{author}{\bibinfo{person}{Steffen Rendle}, \bibinfo{person}{Christoph
  Freudenthaler}, \bibinfo{person}{Zeno Gantner}, {and} \bibinfo{person}{Lars
  Schmidt-Thieme}.} \bibinfo{year}{2009}\natexlab{}.
\newblock \showarticletitle{BPR: Bayesian personalized ranking from implicit
  feedback}. In \bibinfo{booktitle}{\emph{Proceedings of the twenty-fifth
  conference on uncertainty in artificial intelligence}}. AUAI Press,
  \bibinfo{pages}{452--461}.
\newblock


\bibitem[\protect\citeauthoryear{Song, Elkahky, and He}{Song
  et~al\mbox{.}}{2016}]%
        {song2016multi}
\bibfield{author}{\bibinfo{person}{Yang Song}, \bibinfo{person}{Ali~Mamdouh
  Elkahky}, {and} \bibinfo{person}{Xiaodong He}.}
  \bibinfo{year}{2016}\natexlab{}.
\newblock \showarticletitle{Multi-rate deep learning for temporal
  recommendation}. In \bibinfo{booktitle}{\emph{Proceedings of the 39th
  International ACM SIGIR conference on Research and Development in Information
  Retrieval}}. \bibinfo{pages}{909--912}.
\newblock


\bibitem[\protect\citeauthoryear{Wang, Cho, and Wen}{Wang
  et~al\mbox{.}}{2019}]%
        {wang2019attention}
\bibfield{author}{\bibinfo{person}{Tian Wang}, \bibinfo{person}{Kyunghyun Cho},
  {and} \bibinfo{person}{Musen Wen}.} \bibinfo{year}{2019}\natexlab{}.
\newblock \showarticletitle{Attention-based mixture density recurrent networks
  for history-based recommendation}. In \bibinfo{booktitle}{\emph{Proceedings
  of the 1st International Workshop on Deep Learning Practice for
  High-Dimensional Sparse Data}}. \bibinfo{pages}{1--9}.
\newblock


\bibitem[\protect\citeauthoryear{Wang and Fu}{Wang and Fu}{2020}]%
        {wang2020item}
\bibfield{author}{\bibinfo{person}{Tian Wang} {and} \bibinfo{person}{Yuyangzi
  Fu}.} \bibinfo{year}{2020}\natexlab{}.
\newblock \showarticletitle{Item-based Collaborative Filtering with BERT}. In
  \bibinfo{booktitle}{\emph{Proceedings of The 3rd Workshop on e-Commerce and
  NLP}}. \bibinfo{pages}{54--58}.
\newblock


\bibitem[\protect\citeauthoryear{Yi, Yang, Hong, Cheng, Heldt, Kumthekar, Zhao,
  Wei, and Chi}{Yi et~al\mbox{.}}{2019}]%
        {yi2019sampling}
\bibfield{author}{\bibinfo{person}{Xinyang Yi}, \bibinfo{person}{Ji Yang},
  \bibinfo{person}{Lichan Hong}, \bibinfo{person}{Derek~Zhiyuan Cheng},
  \bibinfo{person}{Lukasz Heldt}, \bibinfo{person}{Aditee Kumthekar},
  \bibinfo{person}{Zhe Zhao}, \bibinfo{person}{Li Wei}, {and}
  \bibinfo{person}{Ed Chi}.} \bibinfo{year}{2019}\natexlab{}.
\newblock \showarticletitle{Sampling-bias-corrected neural modeling for large
  corpus item recommendations}. In \bibinfo{booktitle}{\emph{Proceedings of the
  13th ACM Conference on Recommender Systems}}. \bibinfo{pages}{269--277}.
\newblock


\bibitem[\protect\citeauthoryear{Zhou, Zhu, Song, Fan, Zhu, Ma, Yan, Jin, Li,
  and Gai}{Zhou et~al\mbox{.}}{2018}]%
        {zhou2018deep}
\bibfield{author}{\bibinfo{person}{Guorui Zhou}, \bibinfo{person}{Xiaoqiang
  Zhu}, \bibinfo{person}{Chenru Song}, \bibinfo{person}{Ying Fan},
  \bibinfo{person}{Han Zhu}, \bibinfo{person}{Xiao Ma},
  \bibinfo{person}{Yanghui Yan}, \bibinfo{person}{Junqi Jin},
  \bibinfo{person}{Han Li}, {and} \bibinfo{person}{Kun Gai}.}
  \bibinfo{year}{2018}\natexlab{}.
\newblock \showarticletitle{Deep interest network for click-through rate
  prediction}. In \bibinfo{booktitle}{\emph{Proceedings of the 24th ACM SIGKDD
  International Conference on Knowledge Discovery \& Data Mining}}.
  \bibinfo{pages}{1059--1068}.
\newblock


\bibitem[\protect\citeauthoryear{Zhu, Li, Zhang, Li, He, Li, and Gai}{Zhu
  et~al\mbox{.}}{2018}]%
        {zhu2018learning}
\bibfield{author}{\bibinfo{person}{Han Zhu}, \bibinfo{person}{Xiang Li},
  \bibinfo{person}{Pengye Zhang}, \bibinfo{person}{Guozheng Li},
  \bibinfo{person}{Jie He}, \bibinfo{person}{Han Li}, {and}
  \bibinfo{person}{Kun Gai}.} \bibinfo{year}{2018}\natexlab{}.
\newblock \showarticletitle{Learning tree-based deep model for recommender
  systems}. In \bibinfo{booktitle}{\emph{Proceedings of the 24th ACM SIGKDD
  International Conference on Knowledge Discovery \& Data Mining}}.
  \bibinfo{pages}{1079--1088}.
\newblock


\end{thebibliography}

\end{document}